\documentclass[10pt,twocolumn,english,prd,superscriptaddress,nofootinbib,preprintnumbers,showpacs,floatfix]{revtex4-2}

\usepackage[utf8]{inputenc}
\usepackage{amsmath}
\usepackage{amsfonts}
\usepackage{amssymb}
\usepackage[mathlines]{lineno}
\usepackage{graphicx} 
\usepackage[caption=false]{subfig}
\graphicspath{ {figures/} }
\usepackage[usenames,dvipsnames]{xcolor}
\usepackage{hyperref}   
\hypersetup{
	colorlinks=true, 
	citecolor=MidnightBlue,
	linkcolor=MidnightBlue,
	urlcolor=Cyan}
\usepackage{tikz}
\usepackage{verbatim} 
\usepackage{soul} 
\definecolor{purple}{rgb}{1,0,1}
\definecolor{lime}{HTML}{A6CE39} 

\begin{document}
	
\title{Thermodynamics and information recovery of Schwarzschild AdS black holes in conformal Killing gravity}
	
	\author{Yahya Ladghami}
	\email{yahya.ladghami@ump.ac.ma}
	\affiliation{Laboratory of Physics of Matter and Radiation, Mohammed I University, BP 717, Oujda, Morocco}
	\affiliation{Astrophysical and Cosmological Center, BP 717, Oujda, Morocco}
	
	\author{Brahim Asfour} 
	\email{brahim.asfour@ump.ac.ma}
	\affiliation{Laboratory of Physics of Matter and Radiation, Mohammed I University, BP 717, Oujda, Morocco}
	\affiliation{Astrophysical and Cosmological Center, BP 717, Oujda, Morocco}
	
	\author{Francisco S. N. Lobo} \email{fslobo@ciencias.ulisboa.pt}
	\affiliation{Instituto de Astrof\'{i}sica e Ci\^{e}ncias do Espa\c{c}o, Faculdade de Ci\^{e}ncias da Universidade de Lisboa, Edifício C8, Campo Grande, P-1749-016 Lisbon, Portugal}
	\affiliation{Departamento de F\'{i}sica, Faculdade de Ci\^{e}ncias da Universidade de Lisboa, Edif\'{i}cio C8, Campo Grande, P-1749-016 Lisbon, Portugal}
	
	
	\author{Taoufik Ouali} 
	\email{t.ouali@ump.ac.ma}
	\affiliation{Laboratory of Physics of Matter and Radiation, Mohammed I University, BP 717, Oujda, Morocco}
	\affiliation{Astrophysical and Cosmological Center, BP 717, Oujda, Morocco}

\date{\today}
\begin{abstract}
We study Schwarzschild AdS black holes in conformal Killing gravity, focusing on their thermodynamics and information recovery via the island formula. Treating the cosmological constant as pressure and the conformal Killing gravity  parameter as an independent variable, we find that the Bekenstein-Hawking area law holds, while the conformal Killing gravity  parameter dramatically affects phase structure. For a positive conformal Killing gravity  parameter, black holes admit an extremal limit and exhibit Van der Waals-like criticality with first and second order phase transitions; for a negative conformal Killing gravity  parameter, no extremal limit or criticality occurs. 
Using the island prescription, we show that without islands, the entanglement entropy of Hawking radiation grows unboundedly, violating unitarity, while including islands after Page time restores the Page curve, with late-time entropy saturating at twice the Bekenstein-Hawking value. Page time can be expressed in terms of thermodynamic quantities, displaying critical behavior for positive conformal Killing gravity  parameter, whereas in negative conformal Killing gravity small black holes recover information rapidly and large ones more slowly, with pressure reducing Page time. Our results reveal a direct link between black hole thermodynamics, quantum information recovery, and modified gravity.
\end{abstract}
\maketitle


\section{Introduction}

Black holes are fundamental objects in modern theoretical physics, providing a unique laboratory to probe quantum gravity, holography, and unified theories. One of the landmark discoveries in black hole physics is that black holes radiate thermally, a phenomenon known as Hawking radiation \cite{1}. Complementing this, black holes possess an entropy proportional to the area of their event horizons, the Bekenstein–Hawking entropy \cite{2}, and obey four laws analogous to those of thermodynamics \cite{3a,3aa}. These insights established black hole thermodynamics as a cornerstone of gravitational theory and revealed deep connections between gravity and thermodynamic principles. In asymptotically Anti-de Sitter (AdS) spacetimes, black holes exhibit especially rich thermodynamic behavior, including first- and second-order phase transitions \cite{a1,a2,a3,A0,A1,A11,A2,A4,A44,A5}, multicritical points \cite{W,Wu}, Van der Waals–like behavior \cite{L,He,Di}, and critical phenomena \cite{3b,3BB}, making AdS geometries particularly suitable for exploring the interplay between geometry, thermodynamics, and statistical physics.

Despite these advances, Hawking radiation also leads to the long-standing information paradox. As originally observed by Hawking, radiation emitted from a black hole is purely thermal and uncorrelated with the matter that formed it \cite{1}. Consequently, complete evaporation of a black hole transforms an initial pure state into a final mixed state, in apparent violation of the unitarity of quantum mechanics \cite{16}. To reconcile this, Page proposed that if black hole evaporation is unitary, the entanglement entropy of Hawking radiation should follow a characteristic \emph{Page curve}: it increases initially, reaching a maximum at the \emph{Page time}, and subsequently decreases as information is returned to the radiation \cite{17}.  

Two main proposals have been advanced to reproduce the Page curve. The first is the firewall hypothesis \cite{AMPS2013}, which postulates the existence of an energetic barrier at the horizon, destroying infalling observers and thereby violating the equivalence principle. The second, and more widely used in recent studies, is the \emph{island formula} \cite{18}, which incorporates contributions from regions inside the black hole, known as \emph{islands}, in the computation of entanglement entropy. This approach successfully reproduces the Page curve for a variety of black holes across different gravity theories, including Schwarzschild \cite{19,19a}, Reissner–Nordström \cite{20}, noncommutative \cite{21}, BTZ \cite{ytr}, Kaluza–Klein \cite{L2}, and Liouville black holes \cite{Li2021}.

In parallel with these developments, a variety of modified theories of gravity have been proposed to address outstanding challenges in cosmology and fundamental physics beyond general relativity. Among these, \emph{conformal Killing gravity}, introduced by Harada \cite{Harada:2023rqw}, offers a novel geometric framework in which gravitational dynamics are governed by a divergence-free conformal Killing tensor constructed from curvature and matter fields. This formulation can alternatively be expressed as the Einstein field equations supplemented by an additional divergence-free conformal Killing tensor \cite{Mantica:2024cks}. A notable feature of this theory is that the cosmological constant emerges naturally as an integration constant, while the conservation of the energy-momentum tensor, $\nabla_\mu T^{\mu}_{\ \nu} = 0$, arises directly from the field equations rather than being imposed as an independent condition \cite{Harada:2023rqw}.

Moreover, the cosmological evolution equations within this framework inherently include an extra contribution that can be interpreted as an effective dark energy component with a fixed equation-of-state parameter $\omega = - 5/3$, without the need to introduce additional physical fields \cite{Harada:2023rqw}. This feature suggests that the observed late-time acceleration of the Universe may originate from modified gravitational dynamics rather than from a fundamental dark energy sector.

Recent studies have further explored compact objects in this theory. In particular, black bounce solutions have been constructed within conformal Killing gravity by extending static, spherically symmetric configurations and coupling the framework to nonlinear electrodynamics and scalar fields \cite{Mantica:2024cks,Junior:2025izx}. These solutions describe regular spacetimes that interpolate between black hole and wormhole geometries \cite{Junior:2024bb}. Additionally, the thermodynamic and topological properties of static black holes in conformal Killing gravity have been investigated \cite{Chen:2025ut}.

The present work investigates Schwarzschild AdS black holes within conformal Killing gravity, with a particular focus on both thermodynamic properties and quantum information aspects. We explore the effects of the conformal Killing gravity  parameter on the extremal limit, phase structure, and critical behavior of black holes. Furthermore, we examine the black hole information paradox using the island formula, analyzing the Page curve and the interplay between thermodynamic quantities and information recovery. In particular, we study how modifications introduced by conformal Killing gravity influence the onset of the island phase and the scaling of Page time.

This paper is organized as follows. In Sec.~\ref{sec}, we present the Schwarzschild AdS black hole solution in conformal Killing gravity and analyze its thermodynamic properties, including the role of positive and negative conformal Killing gravity  parameters. In Sec.~\ref{secc}, we investigate the information paradox via the island formula, examining the entanglement entropy, Page curve, and Page time for both conformal Killing gravity regimes. Sections~\ref{secpc} and \ref{seceva} explore the correspondence between black hole thermodynamics and information recovery, as well as the evaporation dynamics and scaling of Page time. Finally, in Sec.~\ref{seccc}, we summarize our findings, highlight the interplay between thermodynamics and quantum information, and discuss possible extensions to charged, rotating, or higher-dimensional black holes.  
Throughout this work, we adopt units in which $G = \hbar = c = k_B = 1$.

\section{Black Holes in conformal Killing gravity}\label{sec}

In this section, we briefly review conformal Killing gravity and present the exact Schwarzschild AdS black hole solution within this framework. We then analyze the thermodynamic properties of this black hole and discuss the role played by the conformal Killing gravity  parameter.

\subsection{Black Hole Solution}

In this subsection, we review black hole solutions in conformal Killing gravity. The field equations of this theory are given by \cite{Harada:2023rqw}
\begin{equation}
	\label{fe}
	H_{\alpha\mu\nu} = 8\pi  T_{\alpha\mu\nu},
\end{equation}
where the  tensor $H_{\alpha\mu\nu}$ is  defined as
\begin{eqnarray}
		H_{\alpha\mu\nu} & =  & \nabla_{\alpha}R_{\mu\nu}
		+ \nabla_{\mu}R_{\nu\alpha}
		+ \nabla_{\nu}R_{\alpha\mu} 
		\nonumber \\ 
		&& - \frac{1}{3}
		\left(
		g_{\mu\nu}\partial_{\alpha}
		+ g_{\nu\alpha}\partial_{\mu}
		+ g_{\alpha\mu}\partial_{\nu}
		\right) R ,
\end{eqnarray}
where $R_{\mu\nu}$ is the Ricci tensor and $R$ denotes the Ricci scalar. The tensor $H_{\alpha\mu\nu}$ is totally symmetric in its indices $\alpha$, $\mu$, and $\nu$. It also satisfies the traceless condition \cite{Harada:2023rqw}
\begin{equation}
	\label{ea}
	g^{\mu\nu} H_{\alpha\mu\nu} = 0.
\end{equation}

The quantity $T_{\alpha\mu\nu}$ appearing in Eq.~\eqref{fe} represents the generalized energy-momentum tensor and is defined as \cite{Harada:2023rqw}
\begin{eqnarray}
		T_{\alpha\mu\nu} & =  & \nabla_{\alpha}T_{\mu\nu}
		+ \nabla_{\mu}T_{\nu\alpha}
		+ \nabla_{\nu}T_{\alpha\mu} 
			\nonumber \\ 
		&& - \frac{1}{6}
		\left(
		g_{\mu\nu}\partial_{\alpha}
		+ g_{\nu\alpha}\partial_{\mu}
		+ g_{\alpha\mu}\partial_{\nu}
		\right) T ,
\end{eqnarray}
where $T_{\mu\nu}$ is the usual energy-momentum tensor and $T$ denotes its trace. Similar to the  tensor $H_{\alpha\mu\nu}$, the tensor $T_{\alpha\mu\nu}$ is also totally symmetric and satisfies
\begin{equation}
	\label{eb}
	g^{\mu\nu} T_{\alpha\mu\nu} = 2 \nabla_\mu T_{\alpha}{}^{\mu}.
\end{equation}

The conservation law can be verified directly from the field equations. Multiplying Eq.~\eqref{fe} by $g^{\mu\nu}$ and using the properties given in Eqs.~\eqref{ea} and \eqref{eb}, we obtain
\begin{equation}
	g^{\mu\nu} H_{\alpha\mu\nu} = 16 \pi  \nabla_\mu T_{\alpha}{}^{\mu} = 0,
\end{equation}
which immediately leads to the standard conservation condition $\nabla_\mu T^\mu{}_{\alpha}=0$. This result demonstrates the internal consistency of the theory with respect to matter conservation.

The Schwarzschild AdS black hole solution in conformal Killing gravity is described by the static and spherically symmetric metric \cite{Ghaffari2025}
\begin{equation}
	ds^2 = -f(r)\,dt^2 + \frac{dr^2}{f(r)}
	+ r^2 \left(d\theta^2 + \sin^2\theta\, d\phi^2 \right),
	\label{mtrc}
\end{equation}
with the metric function
\begin{equation}
	\label{ZEA&}
	f(r) = 1 - \frac{2M}{r} + \frac{r^2}{l^2}
	- \frac{\lambda}{5} r^4 .
\end{equation}
Here $M$ represents the black hole mass parameter, $l$ is the AdS curvature radius, and $\lambda$ denotes the conformal Killing gravity  parameter, which characterizes deviations from Einstein gravity due to the conformal Killing gravity contribution. The additional $r^4$ term modifies the asymptotic structure of the spacetime and becomes important at large radial distances. In the limit $\lambda \to 0$, the standard Schwarzschild AdS solution of general relativity is recovered.

The black hole mass can be expressed in terms of the event horizon radius $r_+$ by imposing the horizon condition $f(r_+) = 0$. This leads to
\begin{equation}
	M = \frac{r_+}{2}
	\left( 1 + \frac{r_+^2}{l^2}
	- \frac{\lambda}{5} r_+^4 \right).
\end{equation}

The surface gravity of the black hole is obtained from
\begin{equation}
	\kappa = \frac{f'(r_+)}{2}
	= \frac{1}{2r_+}
	\left( 1 + \frac{3r_+^2}{l^2}
	- \lambda r_+^4 \right),
\end{equation}
which determines the Hawking temperature via
\begin{equation}
	T = \frac{\kappa}{2\pi}
	= \frac{1}{4\pi r_+}
	\left( 1 + \frac{3r_+^2}{l^2}
	- \lambda r_+^4 \right).
\end{equation}
The presence of the conformal Killing gravity  parameter $\lambda$ modifies the temperature-radius relation, indicating that the higher-derivative corrections affect the thermodynamic behavior of the black hole.

The extremal limit corresponds to the condition $\kappa = 0$, where the Hawking temperature vanishes. For black holes in conformal Killing gravity, such an extremal configuration exists only when $\lambda > 0$. Solving $\kappa = 0$ yields the extremal horizon radius
\begin{equation}
	r_e =
	\sqrt{
		\frac{3 + \sqrt{9 + 4 l^4 \lambda}}
		{2 l^2 \lambda}
	}.
\end{equation}
In this extremal state the Hawking temperature vanishes, indicating that the black hole reaches a zero-temperature configuration where further evaporation ceases. For $\lambda \le 0$, the extremal condition cannot be satisfied and therefore no zero-temperature black hole solution exists in that regime.

\subsection{Thermodynamics}

We now investigate the thermodynamic behavior and critical phenomena of Schwarzschild black holes in conformal Killing gravity. The thermodynamic quantities can be obtained by applying the standard laws of black hole thermodynamics. In particular, the entropy follows from integrating the first law of thermodynamics. Using the relation between the mass and temperature, the entropy can be written as
\begin{equation}
	S = \int_{0}^{r_+} \frac{dM}{T}
	= \pi r_+^2 .
\end{equation}
This result shows that the entropy is proportional to the area of the event horizon and therefore satisfies the Bekenstein-Hawking area law. Consequently, the presence of the conformal Killing gravity term does not modify the entropy formula for these black holes, and the standard geometric interpretation of entropy in terms of horizon area remains valid.

In the extended phase space formalism, the cosmological constant is treated as a thermodynamic variable related to the pressure. The AdS radius is therefore expressed in terms of the thermodynamic pressure as
\begin{equation}
	P = \frac{3}{8\pi l^2}.
\end{equation}
With this identification, the black hole mass can be interpreted as the enthalpy of the spacetime rather than the internal energy. Substituting the above relation into the mass expression yields
\begin{equation}
	M = \frac{r_+}{2}
	+ \frac{4}{3} \pi r_+^3 P
	- \frac{\lambda}{10} r_+^5 .
\end{equation}
The last term represents the correction arising from conformal Killing gravity, while the standard Schwarzschild AdS result of general relativity is recovered when $\lambda = 0$.

The first law of black hole thermodynamics in the extended phase space takes the form
\begin{equation}
	dM = T dS + V dP + \Lambda d\lambda ,
\end{equation}
where $V$ is the thermodynamic volume conjugate to the pressure and $\Lambda$ denotes the conjugate quantity associated with the conformal Killing gravity  parameter $\lambda$. We refer to $\Lambda$ as the conformal Killing gravity  potential. From the first law, the thermodynamic quantities can be obtained through partial derivatives of the mass,
\begin{equation}
	\label{th}
	T = \left( \frac{\partial M}{\partial S} \right)_{\lambda,P}
	= \frac{8\pi P r_+^2 - \lambda r_+^4 + 1}
	{4\pi r_+},
\end{equation}
\begin{equation}
	\label{thv}
	V = \left( \frac{\partial M}{\partial P} \right)_{\lambda,S}
	= \frac{4\pi r_+^3}{3},
\end{equation}
and
\begin{equation}
	\Lambda =
	\left( \frac{\partial M}{\partial \lambda} \right)_{S,P}
	= -\frac{r_+^5}{10}.
\end{equation}

These thermodynamic quantities satisfy the generalized Smarr relation
\begin{equation}
	M = 2TS - 2PV - 4\Lambda\lambda .
\end{equation}
The additional term involving $\lambda$ reflects the contribution of conformal Killing gravity to the thermodynamic scaling relation. In the limit $\lambda \to 0$, the standard Smarr relation for the Schwarzschild AdS black hole is recovered.

Using Eqs.~\eqref{th} and \eqref{thv}, the equation of state for the black hole can be written in the form
\begin{equation}
	P = \frac{T}{2r_+}
	- \frac{1}{8\pi r_+^2}
	+ \frac{\lambda r_+^2}{8\pi},
	\qquad
	r_+ = \left( \frac{3V}{4\pi} \right)^{1/3}.
\end{equation}
This equation relates the thermodynamic pressure, temperature, and horizon radius, and plays a role analogous to the equation of state of ordinary thermodynamic systems. The additional term proportional to $\lambda$ introduces a new contribution that significantly modifies the thermodynamic phase structure.

The critical points of the system are determined from the inflection point conditions of the equation of state,
\begin{equation}
	\label{TE}
	\left( \frac{\partial P}{\partial r_+} \right)_{T,\lambda} = 0,
	\qquad
	\left( \frac{\partial^2 P}{\partial r_+^2} \right)_{T,\lambda} = 0 .
\end{equation}
Solving these conditions yields the critical parameters that characterize the phase transition of the black hole system. A real solution exists only when $\lambda > 0$, indicating that critical behavior occurs exclusively for black holes in conformal Killing gravity with a positive conformal Killing gravity  parameter. In contrast, no critical phenomena arise in the Schwarzschild AdS case of general relativity ($\lambda = 0$) or when $\lambda < 0$.

The corresponding critical quantities are
\begin{equation}
	\label{CP}
	r_c = \left( \frac{1}{3\lambda} \right)^{1/4},
	\qquad
	T_c = \frac{2}{\pi}
	\left( \frac{\lambda}{27} \right)^{1/4},
	\qquad
	P_c = \frac{\sqrt{3\lambda}}{4\pi}.
\end{equation}
These results demonstrate that the conformal Killing gravity  parameter plays a central role in determining the thermodynamic phase structure of Schwarzschild AdS black holes. In particular, it controls the existence and location of the critical point, thereby governing the emergence of phase transitions in this gravitational system.

\subsection{Positive conformal Killing gravity}

We now analyze the thermodynamic behavior of Schwarzschild AdS black holes in the regime of positive conformal Killing gravity  parameter, which we refer to as Positive conformal Killing gravity. In this case, the black holes exhibit nontrivial thermodynamic behavior characterized by the appearance of critical points and phase transitions. As discussed in the previous subsection, a real critical point exists only when $\lambda>0$, indicating that the conformal Killing gravity  parameter plays a fundamental role in the phase structure of the system.

To simplify the analysis and highlight the universal features of the thermodynamic behavior, we introduce dimensionless thermodynamic variables normalized by their critical values,
\begin{equation}
	a = \frac{r_+}{r_c},
	\qquad
	p = \frac{P}{P_c},
	\qquad
	t = \frac{T}{T_c}.
\end{equation}
Using these definitions together with the equation of state, the normalized temperature can be expressed as
\begin{equation}
	t = \frac{-a^4 + 6 a^2 p + 3}{8 a}.
\end{equation}
These reduced variables allow us to study the thermodynamic properties independently of the explicit value of the conformal Killing gravity  parameter.

\begin{figure}[t]
	\centering
	\includegraphics[width=\columnwidth]{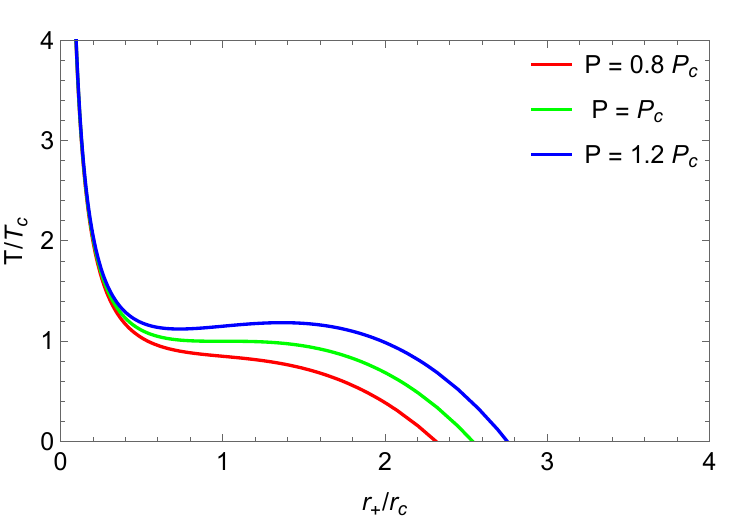}
	\caption{Thermal evolution of Schwarzschild AdS black holes in Positive conformal Killing gravity for different values of the pressure.}
	\label{Tr}
\end{figure}

Fig.~\ref{Tr} illustrates the thermal evolution of Schwarzschild AdS black holes in Positive conformal Killing gravity for different values of the thermodynamic pressure. For small horizon radii, the temperature is relatively high and increases as the event horizon decreases. Conversely, large black holes correspond to lower temperatures, which decrease as the horizon radius increases. 

An interesting feature of this system is the appearance of an extremal configuration in the large black hole regime where the temperature vanishes. In this extremal state, the Hawking temperature becomes zero and the black hole stops emitting Hawking radiation. This behavior differs from the situation in general relativity, where extremal configurations arise only for charged or rotating black holes and are typically associated with the small black hole regime.

The thermodynamic pressure plays a crucial role in determining the phase structure of the system. For pressures below the critical value, a first-order phase transition occurs between small, intermediate, and large black hole phases. At the critical pressure, these phases merge and the system undergoes a second-order phase transition. For pressures above the critical value, the thermodynamic variables vary smoothly and no phase transition occurs.

To further investigate the phase structure, we consider the Gibbs free energy defined by
\begin{equation}
	\label{fff}
	F = M - T S
	= \frac{r_+}{60}
	\left(
	-40 \pi P r_+^2
	+ 9 \lambda r_+^4
	+ 15
	\right).
\end{equation}
The critical value of the Gibbs free energy corresponding to the critical thermodynamic quantities is
\begin{equation}
	F_c = \frac{2}{15}
	\left( \frac{1}{3\lambda} \right)^{1/4}.
\end{equation}
Introducing the normalized Gibbs free energy, we obtain
\begin{equation}
	\frac{F}{F_c}
	= \frac{1}{8}
	a \left( 3 a^4 - 10 a^2 p + 15 \right).
\end{equation}

\begin{figure}[t]
	\centering
	\includegraphics[width=\columnwidth]{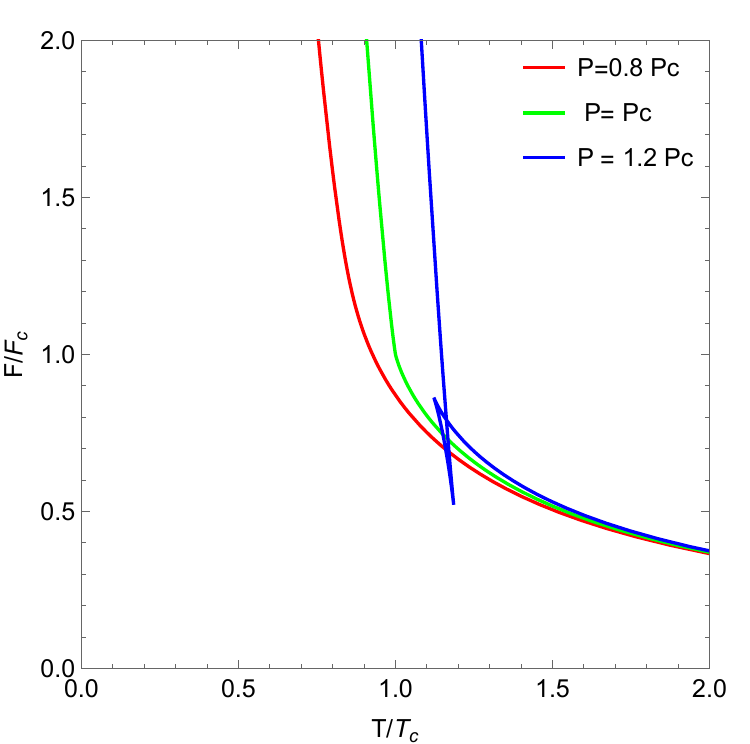}
	\caption{Gibbs free energy as a function of the Hawking temperature for different values of the pressure.}
	\label{ft}
\end{figure}

Fig.~\ref{ft} shows the behavior of the Gibbs free energy as a function of the Hawking temperature for different values of the pressure. The appearance of the characteristic swallow-tail structure indicates the presence of a first-order phase transition. This behavior is analogous to the liquid–gas phase transition of a van der Waals fluid and is a common feature of black holes in the extended phase space.

The local thermodynamic stability of the black holes can be analyzed through the heat capacity at constant pressure and conformal Killing gravity  parameter, which is given by
\begin{equation}
	C =
	T \left( \frac{\partial S}{\partial T} \right)_{P,\lambda}
	=
	\frac{
		2 \pi r_+^2
		\left(
		8 \pi P r_+^2
		- \lambda r_+^4
		+ 1
		\right)
	}{
		8 \pi P r_+^2
		- 3 \lambda r_+^4
		- 1
	}.
\end{equation}

	\begin{figure*}[ht]
		\centering
		\includegraphics[width=0.45\linewidth]{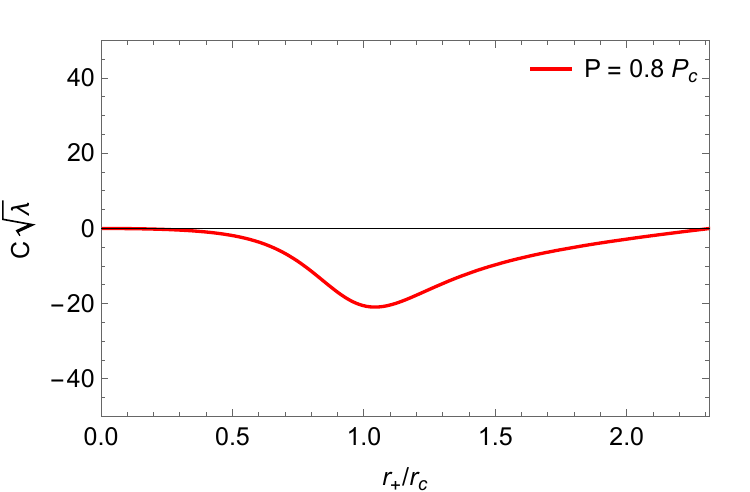}
		\includegraphics[width=0.45\linewidth]{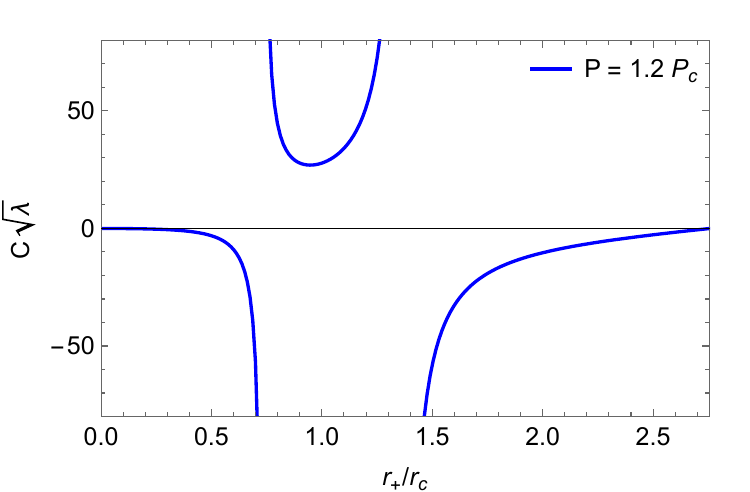}
		\includegraphics[width=0.45\linewidth]{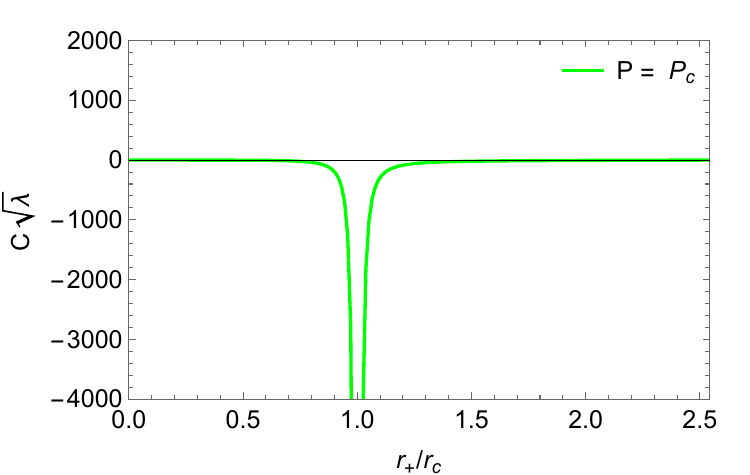}
		\includegraphics[width=0.45\linewidth]{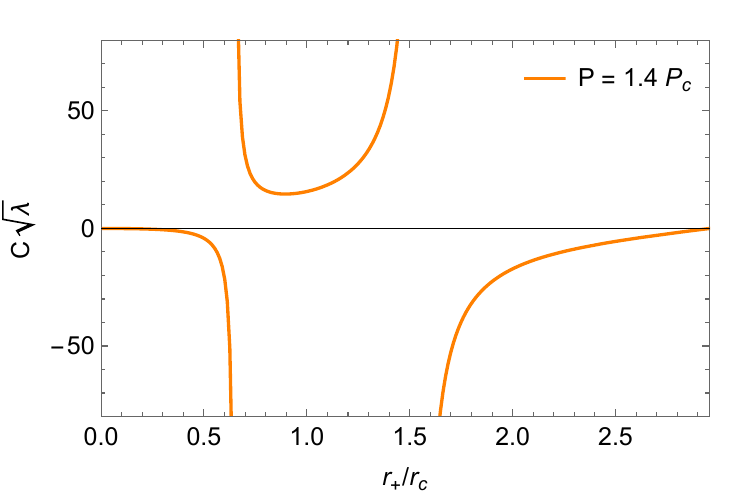}
		\caption{Heat capacity as a function of the event horizon radius for different values of the thermodynamic pressure.}
		\label{hc}
	\end{figure*}

Fig.~\ref{hc} displays the behavior of the heat capacity as a function of the event horizon radius. The sign of the heat capacity determines the local thermodynamic stability of the black hole configurations. Negative values correspond to thermodynamically unstable states, whereas positive values indicate stable configurations.

For pressures below the critical value, the heat capacity is negative over the entire range of horizon radii, indicating that the corresponding black hole configurations are thermodynamically unstable. At the critical pressure, the heat capacity diverges at a specific radius corresponding to the critical point, signaling the onset of critical behavior. For pressures above the critical value, three distinct branches appear. In this regime, the small and large black hole phases remain unstable, while the intermediate black hole phase possesses positive heat capacity and is therefore thermodynamically stable.

\subsection{Negative conformal Killing gravity}

We now consider the case of negative conformal Killing gravity, corresponding to a negative value of the conformal Killing gravity  parameter $\lambda < 0$. In this regime, the thermodynamic behavior of Schwarzschild AdS black holes differs qualitatively from that observed for positive values of the conformal Killing gravity  parameter.

As shown in the previous analysis, the critical conditions do not admit real solutions when $\lambda < 0$. Consequently, no critical phenomena or thermodynamic phase transitions occur in negative conformal Killing gravity. Moreover, unlike the case of positive conformal Killing gravity, Schwarzschild AdS black holes do not possess an extremal configuration when the conformal Killing gravity  parameter is negative. The Hawking temperature therefore never reaches zero, and black holes cannot evolve toward a zero-temperature state.

\begin{figure*}[t]
		\centering
		\includegraphics[width=0.45\linewidth]{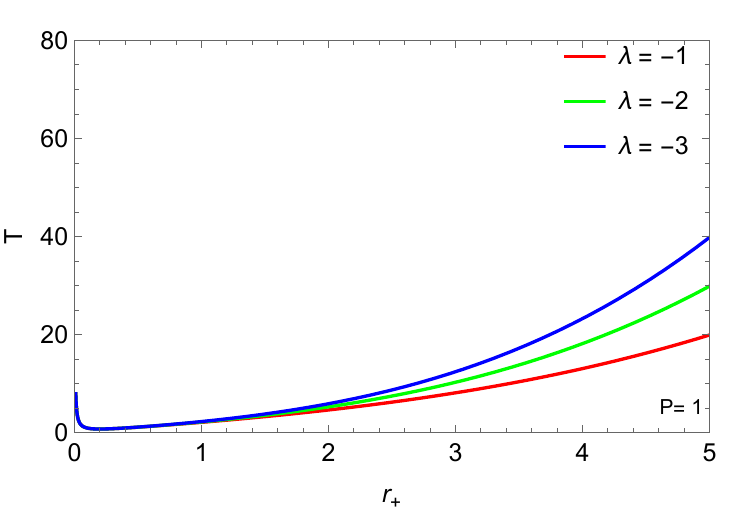}
		\includegraphics[width=0.45\linewidth]{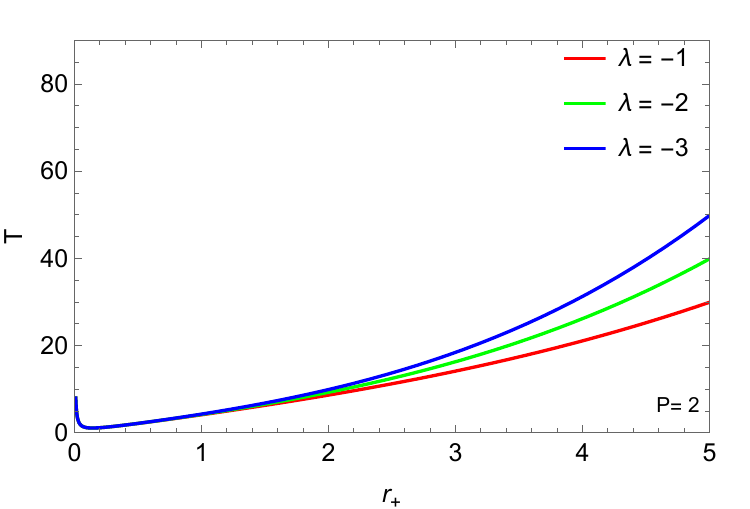}
		\includegraphics[width=0.45\linewidth]{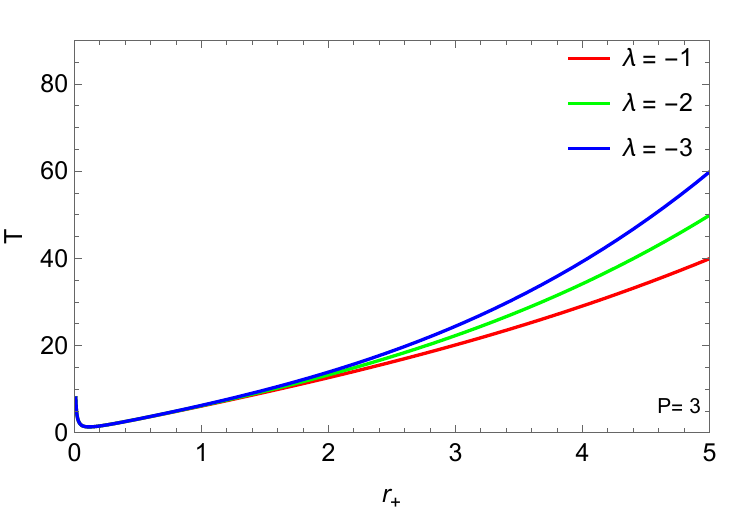}
		\includegraphics[width=0.45\linewidth]{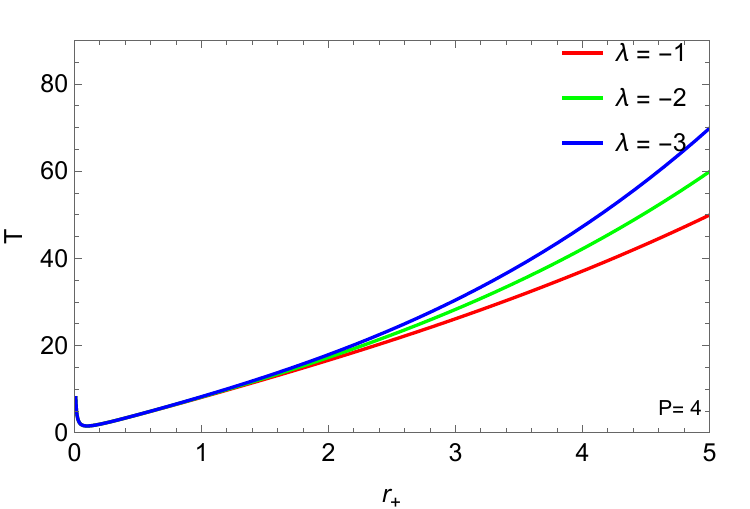}
		\caption{Thermal evolution of Schwarzschild AdS black holes in negative conformal Killing gravity for different values of the thermodynamic pressure.}
		\label{fig:ta}
	\end{figure*}

To investigate the thermal behavior in more detail, we analyze the Hawking temperature as a function of the event horizon radius for different values of the thermodynamic pressure using the temperature expression given in Eq.~\eqref{th}. The corresponding results are displayed in Fig.~\ref{fig:ta}.

Figure~\ref{fig:ta} illustrates the variation of the Hawking temperature for different values of the pressure and the negative conformal Killing gravity  parameter. Two distinct regimes can be identified. The first regime corresponds to small black holes, where the temperature decreases as the event horizon radius increases. The second regime corresponds to large black holes, where the temperature increases with increasing horizon radius.

The thermodynamic pressure mainly affects the overall temperature scale of the system, while the conformal Killing gravity  parameter primarily influences the large black hole regime. In particular, the temperature of large black holes increases as the magnitude of the negative conformal Killing gravity  parameter decreases. This indicates that, in negative conformal Killing gravity, the conformal Killing gravity correction mainly affects the large black hole sector, while the thermodynamic behavior of small black holes remains largely unchanged.

\begin{figure}[t]
	\centering
	\includegraphics[width=\columnwidth]{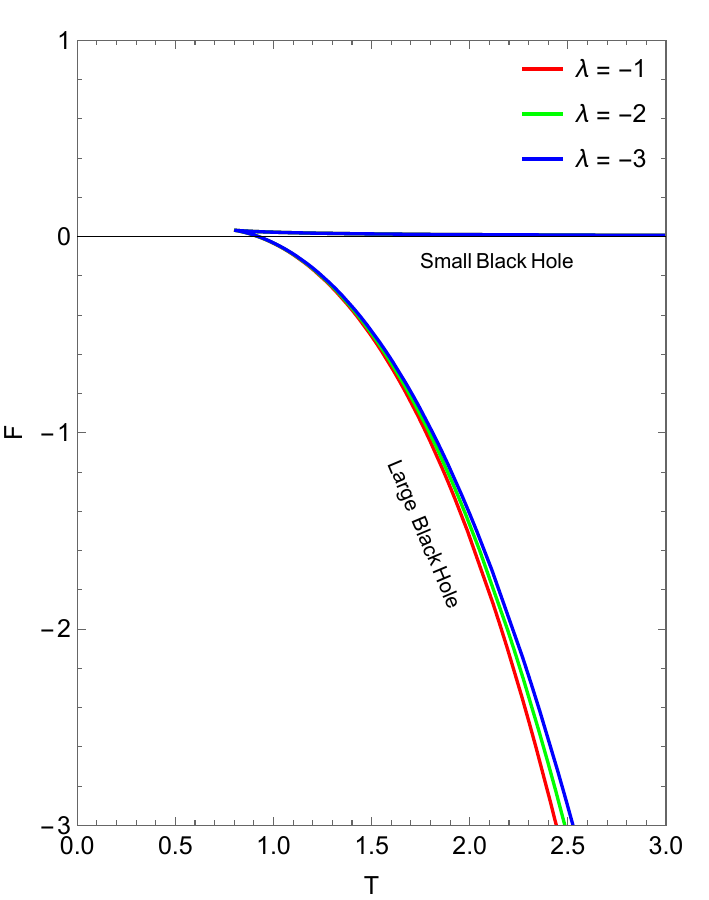}
	\caption{Gibbs free energy as a function of the Hawking temperature for $P=1$ and different values of the negative conformal Killing gravity  parameter.}
	\label{ff}
\end{figure}

Using Eq.~\eqref{fff}, we obtain the Gibbs free energy behavior shown in Fig.~\ref{ff}. The figure presents the Gibbs free energy as a function of the Hawking temperature for different values of the negative conformal Killing gravity  parameter. Two distinct branches appear in the diagram. The horizontal branch corresponds to small black holes and is essentially independent of the conformal Killing gravity  parameter, whereas the vertical branch corresponds to large black holes and depends explicitly on the conformal Killing gravity  parameter.

\begin{figure}[t]
	\centering
	\includegraphics[width=\columnwidth]{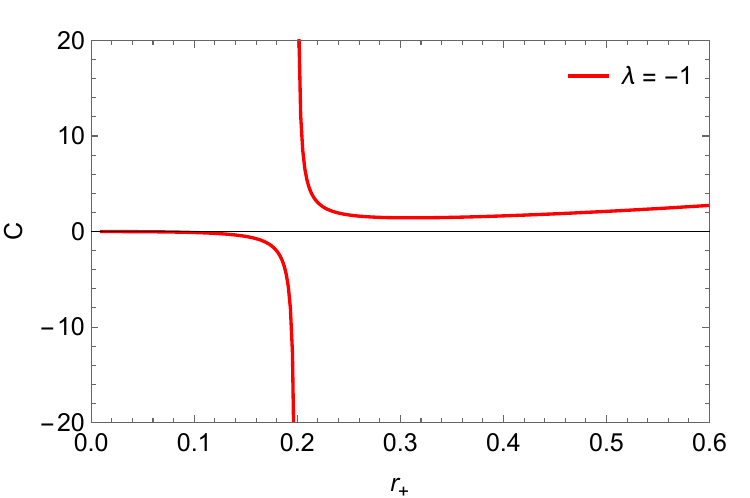}
	\caption{Heat capacity as a function of the event horizon radius for $P=1$ and $\lambda=-1$.}
	\label{hcc}
\end{figure}

From the heat capacity behavior shown in Fig.~\ref{hcc}, we conclude that small black holes are thermodynamically stable, since they possess positive heat capacity. In contrast, large black holes have negative heat capacity and are therefore thermodynamically unstable.

Overall, the conformal Killing gravity  parameter plays a crucial role in determining the thermodynamic behavior of Schwarzschild AdS black holes. For positive values of the conformal Killing gravity  parameter, the system exhibits an extremal limit, van der Waals–like behavior, and critical phenomena. In contrast, when the conformal Killing gravity  parameter is negative, the thermodynamic behavior reduces to that of ordinary Schwarzschild AdS black holes, characterized by a stable small black hole phase and an unstable large black hole phase.

In the following sections, we extend our analysis to explore the implications of conformal Killing gravity for black hole information and the black hole information paradox.

\section{Island Formula}\label{secc}

In this section, we investigate the black hole information paradox and the application of the island formula for Schwarzschild black holes in conformal Killing gravity. In particular, we analyze how the information recovery mechanism operates in both positive and negative conformal Killing gravity regimes.

The black hole information paradox arises from the apparent conflict between Hawking radiation and the principles of quantum mechanics. Hawking's original calculation predicts that black holes radiate thermally, which would imply a loss of information if the black hole evaporates completely. However, recent developments in semiclassical gravity suggest that the paradox can be resolved by including new contributions to the entanglement entropy of Hawking radiation.

A significant advance in this direction is the proposal of the \emph{island formula}, according to which a region inside the black hole contributes to the entanglement entropy of the radiation after the Page time. Before the Page time, the entropy of the radiation is computed without including an island. After the Page time, however, an island region emerges and modifies the entropy calculation.

The entanglement entropy of the radiation including the island contribution is given by \cite{18}
\begin{equation}
	S(R) = \min \left\{ \text{ext} \left[ \frac{\text{Area}(\partial I)}{4} + S_{\text{Bulk}}(R \cup I) \right] \right\},
	\label{TRa}
\end{equation}
where $R$ denotes the region of Hawking radiation outside the black hole, $I$ represents the island region, and $\partial I$ is the boundary of the island. The symbol ``ext'' indicates that the generalized entropy must be extremized with respect to the position of the island boundary $\partial I$. Among all possible extremal configurations, the physical entropy corresponds to the minimal value, as indicated by the symbol ``min''.

To analyze the island prescription for Schwarzschild AdS black holes in conformal Killing gravity, it is convenient to introduce Kruskal coordinates, which provide a regular coordinate system across the event horizon. These coordinates are defined as
\begin{equation}
	\label{Aza}
	U = -e^{-\kappa(t-r_*)}, \qquad 
	V = e^{\kappa(t+r_*)},
\end{equation}
where $\kappa$ denotes the surface gravity of the black hole and $r_*$ is the tortoise coordinate defined by
\begin{equation}
	r_* = \int \frac{1}{f(r)} \, dr.
\end{equation}

The spacetime metric of the Schwarzschild AdS black hole in conformal Killing gravity is given in Eqs.~\eqref{mtrc} and \eqref{ZEA&}. In order to simplify the analysis of the entanglement entropy, we consider the large-distance limit and employ the s-wave approximation. In this approximation, the angular part of the metric is neglected, and the problem effectively reduces to a two-dimensional spacetime described by the radial and temporal coordinates.

Expressed in terms of the Kruskal coordinates, the metric then takes the conformally flat form
\begin{equation}
	ds^2 = W(r)^2 dU dV,
\end{equation}
where the conformal factor $W(r)$ is given by
\begin{equation}
	W(r)^2 = \frac{f(r)}{\kappa^2 e^{2\kappa r_*}}.
\end{equation}

This coordinate system will allow us to evaluate the generalized entropy and determine the location of the island surface in the following subsections.

\subsection{Positive conformal Killing gravity}

We now investigate the information paradox and the island formula for Schwarzschild AdS black holes in the regime of Positive conformal Killing gravity. In this case, extremal black hole configurations exist when $r_+ \ge r_e$. For such extremal black holes, the Hawking temperature vanishes and the black hole does not emit radiation. Consequently, no Hawking radiation is produced and the information paradox does not arise. 

However, for non-extremal black holes satisfying $r_+ < r_e$, the Hawking temperature is nonzero and the black hole undergoes evaporation. In this regime, Hawking radiation carries away information, and the information paradox must be addressed. We therefore begin by calculating the entanglement entropy of Hawking radiation without including the island contribution.

\begin{figure}[t]
	\centering
	\includegraphics[width=\columnwidth]{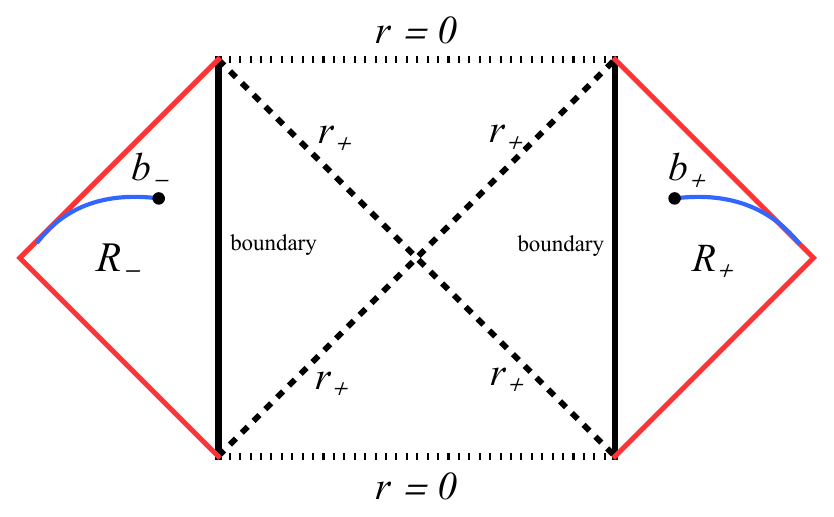}
	\caption{Penrose diagram of Schwarzschild AdS black holes in conformal Killing gravity without an island.}
	\label{wis}
\end{figure}

Fig.~\ref{wis} illustrates the Penrose diagram of the Schwarzschild black hole geometry in conformal Killing gravity without the presence of an island. The Hawking radiation is represented by the region $R$, which consists of two disconnected parts $R_\pm$ located in the right and left wedges of the spacetime. The boundary surfaces of these regions are denoted by $b_\pm$. The corresponding coordinates are $(t_b,b)$ for $b_+$ and $(-t_b + i\beta/2, b)$ for $b_-$, where $\beta = 1/T$ is the inverse Hawking temperature.

We assume that the total system is initially prepared in a pure quantum state at $t=0$. Because the global state is pure, the von Neumann entropy satisfies a complementarity relation between any region and its complement. This property, known as the complementarity principle of von Neumann entropy, implies that the entropy of a subsystem is equal to the entropy of its complementary region.

In our setup, the radiation region $R$ is defined as the union of two semi-infinite intervals,
\begin{equation}
	R = ]-\infty, b_- ] \cup [b_+, +\infty [ ,
\end{equation}
as shown in Fig.~\ref{wis}. The complementary region corresponds to the finite interval $[b_-,b_+]$, which is described by a conformal field theory (CFT). Consequently, the entropy of the CFT in the complementary region is equivalent to the entanglement entropy of Hawking radiation \cite{A12}. The entanglement entropy is therefore given by
\begin{equation}
	S(R) = \frac{c}{3} \log \left[d(b_+, b_-)\right],
\end{equation}
where $c$ is the central charge of the conformal field theory and $d(b_+,b_-)$ denotes the geodesic distance between the endpoints.

The geodesic distance can be written as
\begin{equation}
	\label{AZz}
	d^2(b_+, b_-) = W^2(b)
	\left[ U(b_+) - U(b_-) \right]
	\left[ V(b_-) - V(b_+) \right].
\end{equation}

At large radial distance we have 	$W^2(b)=(1+ \frac{b^2}{l^2} - \frac{\lambda b^4}{5})/(\kappa^2 e^{2\kappa r_*})$.
 Using Eqs.~\eqref{Aza} and \eqref{AZz}, we obtain the entanglement entropy of Hawking radiation
\begin{equation}
	\label{srr}
	S(R) = \frac{c}{3} \log \left( \frac{2\cosh(\kappa t_b)}{\kappa} \right) + \frac{c}{6} \log \left(1+  \frac{b^2}{l^2} - \frac{\lambda b^4}{5} \right).
\end{equation}

At late times ($t_b \to \infty$), this expression reduces to
\begin{equation}
	\label{sr}
	S(R) \approx \frac{c}{3}\kappa t_b
	= \frac{c}{6 r_+}
	\left(1+\frac{3r_+^2}{l^2}-\lambda r_+^4\right)t_b .
\end{equation}

Equation \eqref{sr} shows that the entanglement entropy grows linearly with time and diverges at late times. This behavior violates the expected Page curve, indicating that the radiation remains in a mixed state and does not encode the full information of the initial quantum state. This apparent loss of information leads to the black hole information paradox. To resolve this issue, we must include the island contribution in the entropy calculation after the Page time.

\begin{figure}[t]
	\centering
	\includegraphics[width=\columnwidth]{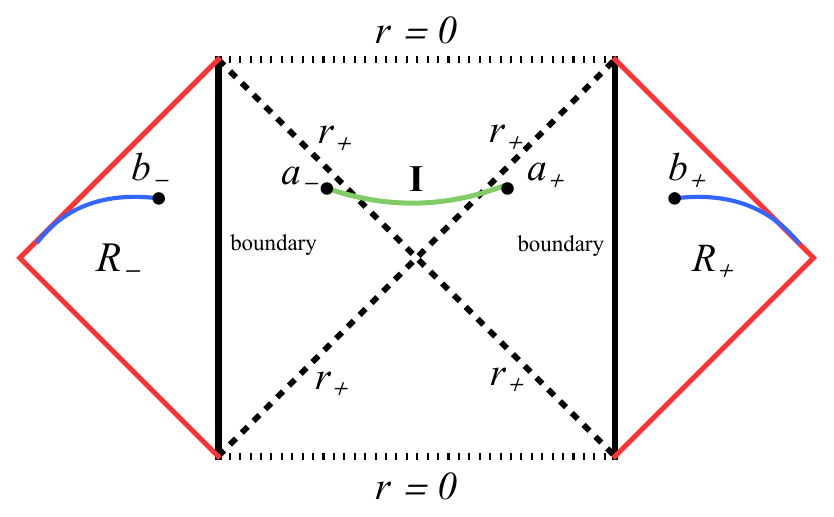}
	\caption{Penrose diagram of Schwarzschild AdS black holes in conformal Killing gravity with an island.}
	\label{is}
\end{figure}

Fig.~\ref{is} shows the Penrose diagram when the island region is included (green curve). The island boundaries are denoted by $a_\pm$, whose coordinates are $(t_a,a)$ for $a_+$ and $(-t_a+i\beta/2,a)$ for $a_-$. The generalized entropy of Hawking radiation including the island is given by \cite{A12}
\begin{eqnarray}
	S_{\text{gen}}(R)
	&=& 2\pi a^2
		\nonumber \\
	&& \hspace{-1.25cm} +\frac{c}{3}\log
	\frac{d(a_+,a_-)d(b_+,b_-)d(a_+,b_+)d(a_-,b_-)}
	{d(a_+,b_-)d(a_-,b_+)} .
\end{eqnarray}

Using the late-time approximation where distances between the left and right wedges are very large, we obtain \cite{19}
\begin{eqnarray}
	d(a_+,a_-)\simeq d(b_+,b_-)\simeq d(a_+,b_-)\simeq d(a_-,b_+) 
		\nonumber \\
	\gg
	d(a_+,b_+)\simeq d(a_-,b_-).
\end{eqnarray}

The generalized entropy then simplifies to
\begin{equation}
	\label{aze}
	S_{\text{gen}}(R)
	=2\pi a^2
	+\frac{c}{3}\log\left[d(a_+,b_+)d(a_-,b_-)\right].
\end{equation}

Using Eqs.~\eqref{Aza} and \eqref{AZz}, we find
\begin{eqnarray}
	\label{azt}
	&d(a_+,b_+)=d(a_-,b_-)
		\nonumber \\
	&=\sqrt{\frac{2\sqrt{f(a)}
			\left(\cosh(\kappa(r_*(a)-r_*(b)))-\cosh(\kappa(t_a-t_b))\right)}
		{\kappa^2}} .
\end{eqnarray}

Substituting this result into Eq.~\eqref{aze}, the generalized entropy becomes
\begin{eqnarray}
	\label{gen}
	S_{\text{gen}}(R)
	&=& 2\pi a^2
	+\frac{c}{3}\log
	\Bigg\{
	\frac{2\sqrt{f(a) f(b)}}{\kappa^2}
	\Big[\cosh(\kappa(r_*(a)\nonumber\\
	&&-r_*(b)))
		- \cosh(\kappa(t_a-t_b))\Big]
	\Big\}.
\end{eqnarray}

The location of the island boundary is determined by extremizing the generalized entropy,
\begin{equation}
	\frac{\partial S_{\text{gen}}}{\partial t_a}=0,
	\qquad
	\frac{\partial S_{\text{gen}}}{\partial a}=0 .
\end{equation}

Solving the first condition yields $t_a=t_b$. The second condition determines the spatial position of the island. Since the island lies close to the event horizon, we write
\[
a=r_+ + \chi^2 r_+ , \qquad \chi \ll 1 .
\]

Using the near-horizon approximations
\begin{equation}
	f(a)\approx2\kappa r_+\chi^2,
	\qquad
	r_*(a)\approx\frac{1}{\kappa}\log\chi ,
\end{equation}
we obtain
\begin{equation}
	a=r_+
	+\left(\frac{c e^{-\kappa r_*(b)}}{12\pi r_+^2}\right)^2 r_+ ,
\end{equation}
with
\begin{equation}
	\chi=\frac{c e^{-\kappa r_*(b)}}{12\pi r_+^2}.
\end{equation}

Substituting these results into the generalized entropy yields
\begin{eqnarray}
	\label{ent}
	S(R)
	&=& 2S+4\chi^2S
		\nonumber \\
	&& \hspace{-1cm} +\frac{c}{6}
	\log\!\left(
	\frac{2r_+ f(b)\left(e^{2\kappa r_*(b)}-4\chi e^{\kappa r_*(b)}+6\chi^2\right)}
	{\kappa^3}
	\right).
\end{eqnarray}

To leading order, the entropy reduces to $S(R)\approx2S$, indicating that the radiation entropy saturates and reproduces the Page curve.

The Page time is obtained by equating the entropies before and after the island contribution,
\begin{equation}
	\label{ptt}
	t_P=\frac{3r_+^2}{cT}
	=\frac{12\pi r_+^3}{8\pi cPr_+^2 - c\lambda r_+^4 + c}.
\end{equation}

This expression shows that the Page time depends explicitly on the thermodynamic parameters of the black hole. Therefore, the evolution of quantum information is closely connected to the thermodynamic properties of the system.

Using Eqs.~\eqref{CP} and \eqref{ptt}, the Page time at the thermodynamic critical point becomes
\begin{equation}
	t_c=\frac{3^{5/4}\pi}{2c\lambda^{3/4}} .
\end{equation}

This result indicates that the critical Page time is determined both by the conformal Killing gravity  parameter $\lambda$, which characterizes the gravitational theory, and by the central charge $c$ of the dual conformal field theory.

\begin{figure}[t]
	\centering
	\includegraphics[width=\columnwidth]{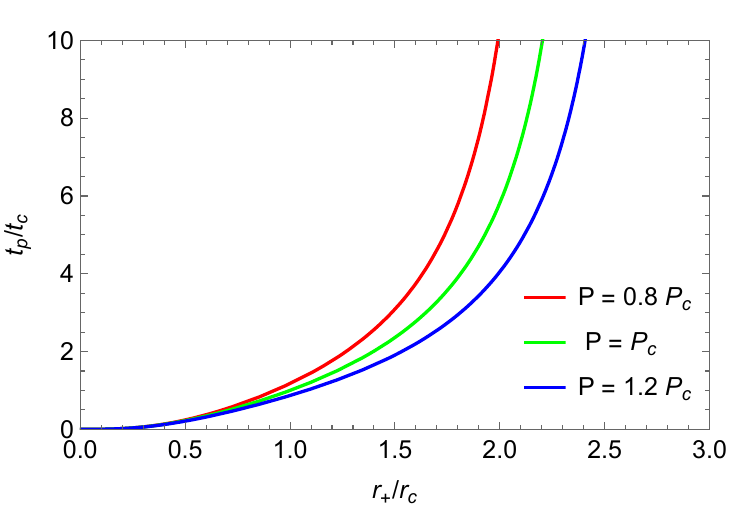}
	\caption{Page time as a function of thermodynamic parameters of the black hole.}
	\label{PTF}
\end{figure}

Fig.~\ref{PTF} shows that the Page time increases with the event horizon radius. Small black holes have a short Page time, implying rapid island formation and faster information recovery. In contrast, large black holes exhibit longer Page times, indicating slower information retrieval. We also observe that increasing the thermodynamic pressure reduces the Page time, particularly for large black holes.

These results highlight a strong interplay between black hole thermodynamics and quantum information. Both the horizon size and the thermodynamic pressure significantly influence the Page time and the dynamics of information recovery in conformal Killing gravity.

\subsection{Negative conformal Killing gravity}

We now study the information paradox and the island formula for Schwarzschild AdS black holes in conformal Killing gravity with a negative conformal Killing gravity  parameter. In this regime, Schwarzschild AdS black holes do not possess an extremal limit. Consequently, the Hawking temperature never vanishes, and the black holes continue to evaporate throughout their evolution until complete disappearance. As a result, the information paradox persists during the entire evaporation process.

We begin by computing the entanglement entropy of Hawking radiation without including the island contribution, following the same procedure as for non-extremal black holes in Positive conformal Killing gravity. In this case, we obtain the same expressions as in Eqs.~\eqref{srr} and \eqref{sr}, with the only difference being the negative sign of the conformal Killing gravity  parameter. As before, the entanglement entropy grows linearly with time at late times and diverges, indicating a violation of the Page curve.

To restore unitarity, we include an island region inside the black hole that emerges after the Page time and contributes to the entropy according to the island formula in Eq.~\eqref{TRa}. Following the same setup as in the previous subsection, we obtain the entanglement entropy of Hawking radiation including the island contribution as in Eq.~\eqref{ent}. To leading order, this entropy can be written in terms of the Bekenstein--Hawking entropy as
\begin{equation}
	S(R) \approx 2S .
\end{equation}

By equating the entanglement entropy of Hawking radiation with and without the island contribution, we obtain the Page time in terms of the thermodynamic quantities of the black hole,
\begin{equation}
	t_P = \frac{12 \pi r_+^3}{8 \pi c P r_+^2 - c \lambda r_+^4 + c}.
	\label{yte}
\end{equation}

In contrast to the case of Positive conformal Killing gravity, the thermodynamic structure of the system does not exhibit critical behavior when the conformal Killing gravity  parameter is negative. Nevertheless, the evolution of Page time still depends on the thermodynamic parameters of the black hole and therefore provides a connection between quantum information and black hole thermodynamics.

\begin{figure}[t]
	\centering
	\includegraphics[width=\columnwidth]{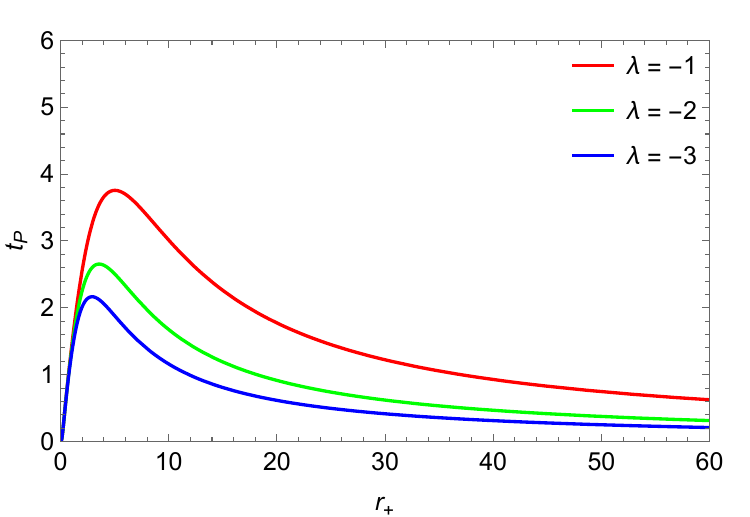}
	\caption{Page time as a function of the event horizon radius and the conformal Killing gravity  parameter for $c=P=1$.}
	\label{ptia}
\end{figure}

Fig.~\ref{ptia} illustrates the evolution of Page time as a function of the black hole event horizon and the conformal Killing gravity  parameter. Two regimes can be identified. The first corresponds to small black holes, where the Page time increases as the horizon radius grows. The second corresponds to large black holes, where the Page time decreases with increasing horizon radius. In addition, the conformal Killing gravity  parameter affects the overall scale of Page time, particularly in the large black hole regime.

These results indicate that quantum information is closely related to the thermodynamic behavior of black holes. The emergence of the island and the recovery of information in Hawking radiation are influenced by the thermodynamic properties of the black hole spacetime. Although negative conformal Killing gravity does not exhibit thermodynamic critical phenomena, the dynamics of quantum information remain sensitive to both the thermodynamic structure and the conformal Killing gravity  parameter. Therefore, the appearance of islands and the associated information recovery process are governed by the underlying thermodynamic properties of Schwarzschild AdS black holes in conformal Killing gravity.

\section{Page Curve and Thermodynamic–Information Correspondence}\label{secpc}

In the previous subsections, we analyzed the entanglement entropy of Hawking radiation for Schwarzschild AdS black holes in conformal Killing gravity by considering both positive and negative values of the conformal Killing gravity  parameter. In both cases, the entropy without the island grows monotonically with time and diverges at late times, indicating a violation of unitary evolution. However, once the island contribution is included, the entropy saturates after the Page time and approaches a constant value proportional to the Bekenstein--Hawking entropy. This behavior reproduces the Page curve and confirms the restoration of unitarity.

The Page curve can be understood as arising from the competition between two possible entropy contributions. The first corresponds to the entropy of Hawking radiation computed without including an island. This contribution increases with time as more radiation escapes from the black hole. The second contribution arises when an island region inside the black hole is included in the entanglement wedge of the radiation. In this case, the generalized entropy receives a dominant contribution from the area of the island boundary, leading to a nearly constant value of the radiation entropy. The physical entropy is determined by the minimal value of these two contributions, which leads to the characteristic Page curve behavior.

An important feature emerging from our analysis is the dependence of the Page time on the thermodynamic parameters of the black hole. In conformal Killing gravity, the Page time depends explicitly on the event horizon radius, the thermodynamic pressure, and the conformal Killing gravity  parameter. Since these quantities also determine the thermodynamic properties of the black hole, such as temperature, stability, and phase structure, the Page time effectively encodes information about the thermodynamic state of the system.

For positive conformal Killing gravity  parameter, the thermodynamic structure includes an extremal limit and a critical behavior similar to that of a Van der Waals fluid. In this regime, the Page time increases with the horizon radius, implying that larger black holes retain information for longer periods before the island forms. This behavior is consistent with the fact that large black holes have lower temperatures and therefore evaporate more slowly. As a result, the process of information recovery is delayed.

In contrast, when the conformal Killing gravity  parameter is negative, the thermodynamic behavior resembles that of ordinary Schwarzschild AdS black holes, with no critical phenomena and no extremal configuration. In this case, the Page time exhibits two distinct regimes. For small black holes, the Page time increases with the horizon radius, while for sufficiently large black holes it begins to decrease. This change reflects the different evaporation dynamics of black holes in this regime and demonstrates that the formation of islands is sensitive to the underlying thermodynamic properties of the spacetime.

These results highlight a direct connection between quantum information and black hole thermodynamics. The emergence of islands, the Page time, and the recovery of information in Hawking radiation are not purely quantum effects but are deeply influenced by the thermodynamic structure of the gravitational theory. Conformal Killing gravity therefore provides a useful framework for exploring how modifications of classical gravitational dynamics affect the quantum information content of evaporating black holes.

Overall, the Page curve in conformal Killing gravity reflects the interplay between semiclassical radiation, generalized entropy, and black hole thermodynamics. The dependence of Page time on the conformal Killing gravity  parameter and other thermodynamic variables suggests that modified gravity theories may provide new insights into the fundamental relationship between spacetime geometry, thermodynamics, and quantum information.

\section{Evaporation Dynamics and Page Time Scaling}\label{seceva}

The analytic expressions obtained for the Page time provide further insight into the evaporation dynamics of Schwarzschild AdS black holes in conformal Killing gravity. Since the Page time represents the moment at which the entanglement entropy of Hawking radiation reaches its maximum and the island contribution becomes dominant, it can be interpreted as the characteristic time scale at which information begins to be recovered from the radiation.

In semiclassical black hole evaporation, the rate of mass loss is governed by the Hawking temperature and the effective radiating area of the horizon. Because both quantities depend on the horizon radius and the thermodynamic parameters of the system, the Page time inherits a nontrivial dependence on the thermodynamic structure of the black hole. In conformal Killing gravity, this dependence becomes particularly transparent through the explicit appearance of the conformal Killing gravity  parameter and thermodynamic pressure in the Page time expressions.

For positive conformal Killing gravity  parameter, the Page time grows monotonically with the event horizon radius. This behavior reflects the slower evaporation rate of larger black holes, which possess lower temperatures and therefore emit Hawking radiation less efficiently. As a result, the entanglement entropy of radiation grows more slowly, and the transition to the island-dominated phase occurs at later times. Consequently, large black holes retain information for longer periods before information recovery begins.

The situation is qualitatively different when the conformal Killing gravity  parameter is negative. In this regime, the Page time exhibits a non-monotonic dependence on the horizon radius. For small black holes, the Page time increases with the horizon size, similarly to the positive conformal Killing gravity case. However, beyond a certain scale, the Page time begins to decrease as the horizon radius increases. This behavior leads to the existence of two distinct evaporation regimes: one associated with small black holes and another associated with large black holes.

Physically, this transition reflects the interplay between horizon geometry and the thermodynamic properties of the spacetime. Small black holes radiate efficiently and therefore reach the Page time relatively quickly. As the horizon grows, the evaporation rate changes in a way that modifies the growth of radiation entropy and the onset of island formation. The resulting two-phase behavior of Page time therefore encodes information about the thermodynamic structure of the underlying gravitational theory.

These observations further reinforce the connection between thermodynamics and quantum information in black hole physics. The Page time, which characterizes the onset of information recovery, is controlled not only by the semiclassical radiation process but also by the thermodynamic parameters governing the black hole solution. In conformal Killing gravity, the conformal Killing gravity  parameter plays a central role in determining this relationship, demonstrating how modifications of gravitational dynamics can influence both the evaporation process and the flow of quantum information.

The scaling behavior of Page time thus provides an additional diagnostic tool for understanding the interplay between gravity, thermodynamics, and quantum information. In particular, the dependence of Page time on the horizon radius and conformal Killing gravity  parameter suggests that modified gravity theories may exhibit distinctive information recovery patterns, potentially offering new insights into the microscopic origin of black hole entropy and the resolution of the information paradox.

\section{Discussion and Conclusions}\label{seccc}

In this work, we investigated Schwarzschild AdS black holes in conformal Killing gravity, focusing on both their thermodynamic properties and their implications for quantum information and the black hole information paradox. In our analysis, the cosmological constant was interpreted as a thermodynamic pressure, while the conformal Killing gravity  parameter was treated as an independent thermodynamic variable. This extended thermodynamic framework allowed us to explore how modifications of gravitational dynamics influence both the phase structure of black holes and the behavior of quantum information encoded in Hawking radiation.

From the thermodynamic perspective, we found that the entropy of Schwarzschild AdS black holes in conformal Killing gravity continues to satisfy the Bekenstein--Hawking area law. Nevertheless, the conformal Killing gravity  parameter plays a decisive role in determining the thermodynamic behavior and phase structure of the system. For positive values of the conformal Killing gravity  parameter, the black holes exhibit an extremal limit and a nontrivial critical phenomenon. In this regime, the thermodynamic behavior resembles that of a Van der Waals fluid, characterized by first- and second-order phase transitions and the existence of a critical point. Such phenomena do not occur either in general relativity or in conformal Killing gravity with negative conformal Killing gravity  parameter. In contrast, when the conformal Killing gravity  parameter is negative, the thermodynamic behavior reduces to that of ordinary Schwarzschild AdS black holes, with no critical behavior and no extremal configuration.

We then investigated the black hole information paradox within conformal Killing gravity using the island formula. For non-extremal black holes, we showed that the entanglement entropy of Hawking radiation diverges at late times when the island contribution is neglected. This behavior violates the Page curve and suggests a breakdown of unitary quantum evolution. However, once the island contribution is included, the entropy saturates after the Page time at a value proportional to twice the Bekenstein--Hawking entropy. In this way, the Page curve is restored, demonstrating that the island mechanism successfully resolves the information paradox even in the presence of modified gravitational dynamics. These results confirm that the island prescription remains robust in conformal Killing gravity.

An important outcome of our study is the connection between information recovery and black hole thermodynamics. In the regime of positive conformal Killing gravity, the Page time increases with the event horizon radius. Small black holes therefore exhibit short Page times, implying rapid island formation and relatively fast information recovery. In contrast, large black holes have longer Page times and recover information more slowly. When the conformal Killing gravity  parameter is negative, the behavior changes qualitatively. In this case, the Page time exhibits two distinct regimes: it increases with the horizon radius for small black holes but decreases for large black holes. These results indicate that the emergence of islands and the recovery of information in Hawking radiation are closely linked to the thermodynamic properties of black holes. More broadly, they suggest that the underlying thermodynamic structure of the gravitational theory plays a fundamental role in determining the dynamics of quantum information.

Our analysis therefore highlights conformal Killing gravity as a useful framework in which modified gravitational dynamics lead to novel thermodynamic behavior that is directly reflected in the Page curve and the information recovery process. The interplay between the conformal Killing gravity  parameter, thermodynamic phase structure, and quantum information provides new insight into how modifications of gravity can influence black hole evaporation and the resolution of the information paradox.

Several interesting directions for future work naturally arise from this study. One important extension would be to analyze charged or rotating black holes in conformal Killing gravity, where additional parameters such as electric charge or angular momentum may further enrich both the thermodynamic phase structure and the island dynamics.

	For charged black holes, the inclusion of a $U(1)$ gauge field introduces an extra thermodynamic degree of freedom and a new length scale.  One may then anticipate a richer phase structure, possibly including re‑entrant phase transitions and multiple critical points, analogous to those observed in charged AdS black holes in Einstein gravity.  The island formula would still be applicable, with the Page time now depending on both the conformal Killing parameter $\lambda$ and the electric charge; near critical points, the Page time could again exhibit critical slowing‑down.  In the rotating case, the addition of angular momentum similarly enlarges the thermodynamic parameter space, leading to a family of Kerr-AdS-like solutions in conformal Killing gravity.  The qualitative picture, the onset of islands and the saturation of the Page curve, is expected to persist, with the Page time encoding the interplay between rotation, the effective cosmological constant, and the modified gravitational dynamics.  A systematic analysis of these extensions would provide a more complete understanding of how modifications of gravity affect black hole thermodynamics and the information paradox.
It would also be worthwhile to investigate higher-dimensional generalizations of conformal Killing gravity, where the structure of the conformal Killing gravity tensor and its thermodynamic consequences may lead to new phenomena.

Another promising direction concerns the holographic interpretation of these results. Since our analysis is performed in an AdS background, it would be interesting to explore the implications of conformal Killing gravity within the framework of the AdS/CFT correspondence, particularly regarding the dual description of island formation and Page curve behavior in the boundary conformal field theory.
From the AdS/CFT perspective, the Schwarzschild--AdS black hole in conformal Killing gravity is dual to a thermal state in a boundary CFT, with the parameter $\lambda$ acting as a deformation of the effective cosmological constant.  At $\lambda=0$ the standard holographic dictionary is recovered, while for $\lambda>0$ the bulk van der Waals criticality maps to a line of first‑order phase transitions and a second‑order critical point in the dual theory.  The island formula and the Page curve remain applicable because the quantum extremal surface is covariantly defined; consequently, the Page time $t_P$ inherits the thermodynamic criticality, exhibiting critical slowing‑down for $\lambda>0$ and monotonic behaviour for $\lambda<0$.  This robustness underscores the universality of the island mechanism in higher‑derivative gravity and directly links the information recovery timescale to the phase structure of the dual field theory.
Finally, it would be valuable to incorporate quantum corrections beyond the semiclassical approximation in order to better understand the interplay between modified gravity, black hole thermodynamics, and the microscopic origin of black hole entropy.

We emphasise that the island formula, being derived from the gravitational path integral, is expected to remain valid beyond the semi‑classical regime.  The qualitative features of the Page curve, namely, the divergence of Hawking entropy without the island, the saturation at twice the Bekenstein-Hawking entropy after the Page time, and the dependence of the Page time on the conformal Killing parameter $\lambda$, are robust predictions that should persist in a full quantum treatment.  The Schwarzschild–AdS solution in conformal Killing gravity is ``Einstein‑like'' and yields the standard Bekenstein–Hawking entropy, which ensures that the leading semi‑classical result is consistent; higher‑derivative corrections (e.g., Wald entropy) would only modify the precise numerical coefficients without altering the overall picture. Moreover, Harada’s theory is ghost‑free on this background, so the semi‑classical approximation is well controlled. Quantum corrections may shift the critical exponents and the exact location of the critical point for $\lambda>0$, but the existence of the van der Waals phase transition and its imprint on the Page curve are expected to survive.

These investigations could provide further insight into the relationship between gravitational dynamics, thermodynamics, and quantum information, and may help clarify the fundamental mechanisms governing information recovery in evaporating black holes.

\acknowledgments{YL gratefully acknowledges the support from the ``PhD-Associate Scholarship -- PASS'' grant provided by the National Center for Scientific and Technical Research in Morocco, under grant number 42 UMP2023.
FSNL acknowledges support from the Funda\c{c}\~{a}o para a Ci\^{e}ncia e a Tecnologia (FCT) Scientific Employment Stimulus contract with reference CEECINST/00032/2018, and funding through the grants UID/04434/2025, and PTDC/FIS-AST/0054/2021.}



\end{document}